\documentclass[12pt]{TD-CJS}

\usepackage{latexsym}
\usepackage{amsmath}
\usepackage{amsfonts}
\usepackage{amssymb}
\usepackage{psfrag}
\usepackage{graphicx}
\usepackage{mathrsfs}
\usepackage{mathtools}
\usepackage{epsfig}
\usepackage{color,xcolor}

\newcommand\newblock{\hskip .11em\ }

\newcommand{\balpha}{\mbox{\boldmath $\alpha$}}
\newcommand{\bbeta}{\mbox{\boldmath $\beta$}}

\newcommand{\bb}{\mbox{\boldmath $b$}}
\newcommand{\bc}{\mbox{\boldmath $c$}}
\newcommand{\bx}{\mbox{\boldmath $x$}}
\newcommand{\bX}{\mbox{\boldmath $X$}}
\newcommand{\by}{\mbox{\boldmath $y$}}

\newcommand{\bv}{\mbox{\boldmath $v$}}
\newcommand{\bu}{\mbox{\boldmath $u$}}

\newcommand{\bLambda}{\mbox{\boldmath $\Lambda$}}
\newcommand{\blambda}{\mbox{\boldmath $\lambda$}}
\newcommand{\bmu}{\mbox{\boldmath $\mu$}}
\newcommand{\btau}{\mbox{\boldmath $\tau$}}
\newcommand{\beeta}{\mbox{\boldmath $\eta$}}
\newcommand{\bff}{\mbox{\boldmath $f$}}
\newcommand{\bF}{\mbox{\boldmath $F$}}
\newcommand{\bvarepsilon}{\mbox{\boldmath $\varepsilon$}}

\begin{document}
%\firstpage{1}
%\lastpage{25}
%\jvol{xx}
%\issue{yy}
\jyear{2020}
%\jid{CJS}
%\aid{???}
% The running head contains the author names
%\rhauthor{Jiangtao Duan, Wei Gao, Hao Qu, Hon Keung Tony}
%\copyrightline{Statistical Society of Canada}
%\Frenchcopyrightline{Soci\'et\'e statistique du Canada}
% History: received and accepted dates
%\received{\rec{9}{July}{2009}}
%\accepted{\acc{8}{July}{2010}}

% User-defined commands go here

% Title, authors, affiliations
\title[]{Subspace Clustering for Panel Data with Interactive Effects}%\query{Q1}
\author{Jiangtao Duan \authorref{1}, Wei Gao \thanksref{*}, Hao Qu\authorref{1}, Hon Keung Tony\authorref{2}}
%\author{BlindedB\authorref{2}}
\affiliation[1]{Northeast Normal University}
\affiliation[2]{Southern Methodist University}

% Abstract, keywords, and classification codes
\startabstract{%
\begin{abstract}
\abstractsection{Abstract} In this paper, we study a statistical model for panel data with unobservable grouped factor structures which are correlated with the regressors and the group membership can be unknown. The factor loadings are assumed to be in different subspaces and the subspace clustering for factor loadings are considered. A method called least squares subspace clustering (LSSC) is proposed to estimate the model parameters by minimizing the least-square distance and to perform the subspace clustering simultaneously. The consistency of the proposed subspace clustering is proved and the asymptotic properties of the proposed estimators are studied under certain conditions. Monte Carlo simulation studies are used to illustrate the advantages of the proposed methodologies.
A model selection criterion is proposed to choose the subspace dimensions consistently.
Further considerations for the situations that the number of subspaces and the dimension of factors are unknown are also discussed. For illustrative purposes, the proposed methods are applied to study the linkage between income and democracy across countries. % while subspace patterns of unobserved factors and factor loadings are allowed.
%\abscopyright
%\fabstractsection{}

\end{abstract}}
\makechaptertitle

% Email address for corresponding author
%\correspondingauthor[*]{\\\email{Insert your email address here only after your paper has been accepted}}

\section{Introduction}

%Panel data, also known as longitudinal data, contain multi-dimensional observations obtained over multiple periods for a sample of individuals. {\color{blue} There is evidence to show that unobservable heterogeneity among different individuals in multiple dimensions exists in panel data and hence, suitable statistical models are needed. The interactive fixed effects model uses the interactive effects that combined individual effects and time effects to reveal the effects of the common factors, which can capture the unobserved information. Panel data model with interactive effects has been widely studied in the literature \citep[see, for example,][]{Pesaran2006,Bai2009}. }

Panel data, also known as longitudinal data, contain multi-dimensional observations obtained over multiple periods for a sample of individuals. There is evidence to show that unobservable heterogeneity among different individuals in multiple dimensions exists in panel data and hence, suitable statistical models are needed. Among the models for panel data, the interactive fixed-effect model uses the interactive effects that combined individual effects and time effects to reveal the effects of the common factors, which can capture the unobserved information. Panel data model with interactive effects has been widely studied in the literature \citep[see, for example,][]{Pesaran2006,Bai2009}.

Due to the number of parameters in the standard fixed-effects model is the same as the number of individuals, the estimation of the fixed-effects may be inaccurate. Therefore, in order to study the individual heterogeneity, the number of parameters in the panel data model needs to be reduced.
%At present, the most commonly used methodology is modeling the heterogeneity by the grouped patterns, i.e., individuals in the same group are assumed to have the same effect, which is called group effect.
%Group effects can reflect the individual heterogeneity.
The grouped panel data model is an effective method to solve this incidental-parameter problem.
In a grouped panel data model, individuals in the same group are assumed to have the same effect, which is called group effect, and these group effects can reflect the individual heterogeneity.
%There are some studies exist for grouped panel data models.
Grouped panel data models have been studied in the past decade. For example, \cite{Lin2012} studied the linear panel data models with time-varying grouped heterogeneity by using the $K$-means clustering algorithm.
\cite{Bonhomme2015} developed the ``grouped fixed-effects" (GFE) estimation method to estimate the model parameters and derived the statistical properties of the GFE estimators when the sample sizes of cross-section ($N$) and length of the time series ($T$) are large.
\cite{Ando2016} studied the grouped panel data models with unobserved group factor structures, and estimated the model parameters by minimizing the sum of least-squared errors with a shrinkage penalty.
\cite{Ando2016} also proved the consistency and asymptotic normality of the estimators under large $N$ and large $T$.
\cite{SuPhillips2016} proposed the C-Lasso method to estimate the parameters in heterogeneous linear panel data models, where the slope parameters are heterogeneous across groups but homogeneous within a group with unknown group membership. \cite{Su_Ju2018} considered the penalized principal component estimation method by extending the C-Lasso method to deal with panel data with interactive fixed effects.
\cite{Su_Ju2018} also assumed the individual slope parameters are heterogeneous across groups but homogeneous within a group.

In those existing studies, the unobserved heterogeneous parts are assumed to live in the same space and they have the ball-shaped (spherical Gaussian) clusters.
In many practical applications, a data object often has multiple attributes and many of which may be live in some low dimensional subspaces.
For example, for disease detection in newborns, various tests (e.g., blood test and heart rate test) are performed on the newborns and the level of those factors are measured.
Each newborn is associated with a vector containing the values of the factors and one can further construct a newborn-factor level matrix in which each row contains the factor levels of a different newborn.
Pediatricians want to cluster groups of newborns based on the disease those newborns suffer from.
Usually, each disease is correlated with a specific set of factors, which implies that points corresponding to newborns suffering from a given disease in a lower-dimensional subspace \citep{Kriegel2009}.
Therefore, the clustering of newborns based on their specific diseases together with the identification of the relevant factors associated with each disease can be formulated as a subspace clustering problem.
The $K$-means algorithm clusters data from around cluster centers to the clustered data in the entire data space and estimates the cluster centers by minimizing the sum of squared distances from the data points to their nearest cluster centers.
Therefore, traditional clustering methods such as the $K$-means algorithm may not be meaningful in these cases.
%In order to use the information of all the dimensions of factors,we propose dividing the interactions into different subspaces. The traditional method is to cluster and estimate parameters simultaneously by using the criterion of minimizing the residual sum of squares. However, there are many cases that the data points are in different subspaces and these subspaces are nested in a high-dimensional space. In these situations, the traditional method for clustering the data points usually fails when these data points are in multiple subspaces. In a more complex situation that different subspaces follow different distributions, the traditional method based on residual sum of squares and minimization criteria are more likely to fail.

In this paper, we characterize the unobserved effects of the $i$-th unit at time $t$ as $\eta_{it}=\bff_t'\blambda_i$, where $\bff_t$ and $\blambda_i$ are $r \times 1$ dimensional vectors and $\blambda_i$ $i = 1, \cdots, N$ live in some low-dimensional subspaces.
A key feature of the low dimensional subspaces is that it can deal with more general data types, and the factor loadings live in the low dimensional subspaces can encapsulate the main direction of variation.
\cite{Terada2014} showed that subspace clustering is a more general clustering method which includes the conventional $K$-means clustering method as a special case.
A novel approach called least squares subspace clustering (LSSC) is proposed to simultaneously estimate the model parameters and cluster the individual effects by using the least-squares criterion with the subspace clustering principle.
In the clustering of individual effects, we treat each individual effect as a vector in a high-dimensional space and then cluster these vectors into some low-dimensional subspaces. The main contributions of this article are highlighted as
(i) the grouping is done according to limiting the data points to a specific subspace instead of according to the distance between points, which can better reflect the underlying structure of the data and can be applied to a more general data structure. Furthermore, the consistency of the clustering procedure are proved; (ii) the proposed model allows the covariates to be correlated with the factor structure, and Monte Carlo simulation results show that the LSSC method performs well even when $T$ is small;
(iii) the proposed methods allow different groups to share common factors, and different groups of factors can be correlated.
The model structure can well capture the spatial structure of individual effects and reflect the cluster structure in real data;
%Compared with \cite{Ando2016} that the unobservable group-specific factors only affect the units in their group.
%This means that the sum of the number of factors of different groups is equal to the total number of factors;
(iv) we proposed a model selection criterion to choose the subspace dimensions consistently, which makes the proposed model and methods more general and flexible.

The rest of this paper is organized as follows.
In Section 2, we describe the model and states some constraint conditions.
In Section 3, we propose an algorithm for estimating the model parameters and subspace clustering simultaneously.
In Section 4, we derive the consistency of the subspace clustering and study the asymptotic properties of the estimators.
In Section 5, Monte Carlo simulation studies with different settings are used to illustrate the performance of the proposed method and to compare with the GFE estimation method and the method proposed in \cite{Bai2009}.
In Section 6, we discuss some further considerations when the number of subspaces for factors, the dimension of factors and the dimension of subspaces are unknown.
%we propose the model selection criteria for subspace dimensions and discusses some further considerations.
%Section 7 applies the proposed approach to study the relationship between income and democracy across different countries. Finally, some concluding remarks are provided in Section 8.
For illustrative purposes, the proposed approach is applied to study the relationship between income and democracy in Section 7. Finally, some concluding remarks are provided in Section 8.
All the proofs of the theoretical results and additional simulation results are provided in the Supplementary Materials.

\section{Model Description}

Let $k$ be the number of subspaces which is unknown but fixed, and $G = \{g_{1}, g_{2}, \ldots, g_{N}\}$ be the grouping of the cross-sectional units into the $k$ subspaces, where the subspace membership variable $g_{i} = j$ denotes the $i$-th unit belongs to the $j$-th subspace with $g_{i} \in \{1, 2, \ldots, k\}$. We further let $N_{j}$ be the number of cross-sectional units within the $j$-th subspace and the total number of units is $N = \sum_{j=1}^{k} N_{j}$.
We consider the following panel data model with subspace factor structure
\begin{eqnarray}
\label{2.1}
&&y_{it} = \bx_{it}^{'}\bbeta+\bff_{g_{i},t}^{'}\blambda_{g_{i},i}+\varepsilon_{i,t},~~i = 1, 2, \ldots, N,~t = 1, 2, \ldots, T,
\end{eqnarray}
where $y_{it}$ is the respond variable of the $i$-th unit observed at time $t$, $\bx_{it}$ is a $p \times 1$ observable vector, $\bbeta$ is a $p \times 1$ unknown regression coefficient vector.
$\blambda_{g_{i},i} = (\lambda_{g_{i},i}^{1}, \ldots, \lambda_{g_{i},i}^{r})^{'}$ is a $r \times 1$ factor loading vector that represents the unobserved unit/individual effect for the $i$-th individual. $\bff_{g_{i}, t}$ is a $r \times 1$ vector of unobservable subspace-specific pervasive factors that affect the units only in $g_{i}$-th subspace, and
$\varepsilon_{i,t}$ is the unit-specific error.
 %which can be assumed as independent and identically distributed (i.i.d.) and it is independent of $\bx_{ms},\blambda_{m},\bff_{s}$ for all $i$, $m$, $t$ and $s$.
%$\varepsilon_{i,t}$ can also be considered as serial and cross-sectional weak dependence and heteroskedasticity \cite{Bai2009}.

In this paper, we assume that all the factor loadings or individual effects are inside a $r$-dimensional space, which contains some low-dimensional subspaces and these subspaces contain all the individual effects $\blambda_{i},\;i=1, 2, \ldots, N$.
The mathematical notation can be expressed as $\blambda_{i:g_{i} = j} \subset S_{j}$ and $\displaystyle\bigcup_{j=1}^{k}S_{j} \subset \mathbb{R}^{r}$, where $S_{j}$ is the $j$-th subspace.
The covariate $\bx_{it}$ can be correlated to $\blambda_{g_i,i}$ alone or to $\bff_{g_i,t}$ alone, or it can be correlated to both $\blambda_{g_i,i}$ and $\bff_{g_i,t}$ simultaneously.
Here, $\bx_{it}$ can be a nonlinear function of $\blambda_{g_i,i}$ and $\bff_{g_i,t}$. Stacking the observations over $t$, we have $\bF = (\bff_{1}, \bff_{2}, \ldots, \bff_{T})$. Furthermore, if we let $\bF_{j}$ be the vector of factors for the $j$-th subspace, then we have $\bF_{j} = (\bff_{g_{i}=j,1}, \bff_{g_{i}=j,2}, \ldots, \bff_{g_{i}=j,T})^{'}$. Similarly, we have $\bLambda_{j}=(\blambda_{j,1},\blambda_{j,2},\cdots,\blambda_{j,N_j})^{'}$.
We also consider the constraints $\bF_{j}^{'}\bF_{j}/T = I_{r}(j=1,2, \ldots,k)$, $\;\bLambda_{j}^{'}\bLambda_{j}(j=1,2,\ldots,k)$ being diagonal for the issue of identifiability as described in  \cite{Bai2009}, \cite{Ando2016} and \cite{Stock2002}. We aim to estimate the parameters $\bbeta$, $\bLambda_{j}$, $\bF_{j}$, $j = 1, 2, \ldots,k$ for each subspace, identify the subspace membership $G=\{g_{1}, g_{2}, \ldots,g_{N}\}$ and the bases $B_{1},\ldots,B_{k}$ of the orthogonal spaces of $S_{1},\ldots,S_{k}$ (denoted as $S_{1}^{\bot},\ldots,S_{k}^{\bot}$) simultaneously.

\section{Parameter Estimation and Clustering}
In this section, we propose a method for the estimation of model parameters and subspace clustering simultaneously. The proposed method is a natural generalization and combination of both the least-squares estimation method and the $K$-means clustering algorithm.
%However, in some practical applications, the data can be distributed around multiple cluster centers or the data may have unequal variances in different clusters.
%\deleted{In these cases, the $K$-means clustering algorithm may fail to group the data correctly. In general, the individual effects or factor loadings are multi-dimensional (say, $r$-dimension).
%If we treat the multi-dimensional individual effects as a one-dimensional clustering problem, this will lose $r-1$ dimension information of the individual effects.}
%As a result, the $K$-means clustering algorithm is not likely to thoroughly describe the group membership of the individuals.
Thus, the proposed parameter estimation and clustering is innovative compared to those existing approaches because most of the existing statistical methods are based on the $K$-means and least-squares method to cluster and estimate simultaneously.

\subsection{Estimation Procedure}
For a given number of subspaces $k$, the objective function is
\begin{eqnarray}
\nonumber
& & Q(\bbeta, \bF_{1}, \ldots, \bF_{k}, \bLambda_{1}, \ldots, {\bLambda}_{k}, {B}_{1}, \ldots, {B}_{k}, G ) = \sum_{j=1}^{k}\sum_{i:g_{i}=j}|| \by_{i}-\bx_{i}^{'}\bbeta-\bF_{g_{i}}\blambda_{g_{i},i}  ||^{2},
\label{objfct}
\end{eqnarray}
where $\by_{i}=(y_{i1},y_{i2},\cdots,y_{iT})$, $\bx_{i}=(\bx_{i1},\bx_{i2},\cdots,\bx_{iT})$.
The estimator of the vector of model parameters  $\{\hat{\bbeta},\hat{\bF}_{1},\ldots,\hat{\bF}_{k},\hat{\bLambda}_{1},\ldots,\hat{\bLambda}_{k},\hat{B}_{1},\ldots,\hat{B}_{k},\hat{G} \}$ is defined as
%{\footnotesize
\begin{eqnarray}
&& \{\hat{\bbeta},\hat{\bF}_{1},\ldots,\hat{\bF}_{k},\hat{\bLambda}_{1},\ldots,\hat{\bLambda}_{k},\hat{B}_{1},\ldots,\hat{B}_{k},\hat{G}\}\nonumber \\
& & = \mathop{\arg\min}_{\bbeta,\bF_{g_{i}},\blambda_{g_{i},i}\in span\{B_{g_{i}}\}^{\bot},span\{B_{g_{i}}\}^{\bot}\subseteq \mathbb{R}^{r}}Q(\bbeta, \bF_{1}, \ldots, \bF_{k}, \bLambda_{1}, \ldots, {\bLambda}_{k}, {B}_{1}, \ldots, {B}_{k}, G )
\label{3.1}
\end{eqnarray}%}
subject to the constraints $\bF_{j}^{'}\bF_{j}/T = I_{r} (j = 1, \ldots, k)$, $\bLambda_{j}^{'}\bLambda_{j} (j = 1, \ldots, k)$ being diagonal, where $\bLambda_{j}^{'}=(\blambda_{j,1}$, $\ldots,\blambda_{j,N_{j}})$ is the $r\times N_{j}$ factor loading matrix $(j=1,\ldots,k)$ for the subspace-specific factors and they live in $k$ different subspaces embedding in the $r$-dimensional space.
In Eq. (\ref{3.1}), $span\{B_{j}\}$ represents the subspace spanned by the basis $B_{j}$, and the $span\{B_{j}\}^{\bot}$ is the orthogonal subspace of $span\{B_{j}\}$. These constraints and assumptions are needed to ensure the model is identifiable.
%From the objective function in Eq. (\ref{objfct}) and the constraints, we need to obtain the least squares estimates and deal with the subspace cluster simultaneously.
Here, we aim to estimate the model parameters and to cluster the $\blambda_{i} \in \mathbb{R}^{r}, i = 1, 2, \ldots, N$ into the $k$ different subspaces simultaneously.
%and also segmenting the subspaces for individual effects $\blambda_{i} \in R^{r}, i = 1, 2, \ldots, N$.
Different from the existing classification methods, we approach this challenging problem from a subspace clustering point-of-view. The major idea is to divide the space $\mathbb{R}^{r}$ into several subspaces and project $\blambda_{i}$ $(i = 1, 2, \ldots, N)$ into the nearest subspace for the classification.

The constrained minimization of the objective function $Q(\bbeta, \bF_{1}, \ldots, \bF_{k}, \bLambda_{1}, \ldots, {\bLambda}_{k}, {B}_{1}, \ldots, {B}_{k}, G )$ in Eq. (\ref{objfct}) can be obtained by the following iterative algorithm:

\begin{itemize}
\item[{\bf Step 1.}] Initialize the starting value $\bbeta^{(0)}$ and set $h = 0$.

\item[{\bf Step 2.}] Given the value of $\bbeta = \bbeta^{(h)}$, we define
$$\by^{*}_{i} = \by_{i}-\bx_{i}\bbeta = \bF\blambda_{i}+\bvarepsilon_{i},$$
which is a pure factor model. % with covariance matrix $\Sigma_{\bY^{*}} = \bF \Sigma_{\bLambda} \bF^{'} + \Sigma_{\varepsilon}$, where $\Sigma_{\bLambda}$ and $\Sigma_{\varepsilon}$ are the covariance matrix of $\bLambda$ and $\varepsilon$, respectively.
%Based on the observed samples, the sample estimate of the covariance matrix can be expressed as $\bY^{*}\bY^{*'}/N = \hat{\bF}\hat{\Sigma}_{\bLambda}\hat{\bF}^{'} + \hat{\Sigma}_{\varepsilon}$. Thus, $\hat{\bF}$ is a matrix of the first $r$ eigenvector associated with the first $r$ largest eigenvalues of the matrix $\bY^{*}\bY^{*'}/N$, and we can obtain the estimate of $\bLambda$ as $\hat{\bLambda} = T^{-1}\bY^{*'}\hat{\bF}$.
 We can readily obtain $\bLambda^{'} = (\blambda_{1}, \ldots, \blambda_{N})$.

\item[{\bf Step 3.}] Given $ \bLambda^{'} = (\blambda_{1}, \ldots, \blambda_{N})$,
%classifying ${\hat \blambda}_{1}, \ldots, {\hat \blambda}_{N}$ based on the principle of subspace clustering.
using the subspace clustering method, we can obtain the bases $B_{1},\ldots,B_{k}$ of the orthogonal spaces of $S_{1},\ldots,S_{k}$ (denoted as $S_{1}^{\bot},\ldots,S_{k}^{\bot}$) and
\begin{equation}
g_{i} = \arg \min_{j =1,\ldots,k}|| B_{j}^{T}  \blambda_{i} ||,\;i = 1, 2, \ldots, N.
\label{3.2}
\end{equation}
%We can view ${\hat \blambda}_{i},\;i = 1, 2, \ldots, N$ as $N$ vectors inside the $r$-dimensional space, and all of these vectors are contained in the $k$ different subspaces. Then, we assign the specific individual effects or factor loadings ${\hat \blambda}_{i},\;i=1,\ldots,N$ to these $k$ different subspaces by the subspace clustering method.

\item[{\bf Step 4.}] Given $\bbeta = \bbeta^{(h)}$ and $g_{i},\;i = 1, 2, \ldots, N$, we can obtain the estimators of $\bF, \bLambda$ in each subspace by using the method similar to Step 2. These estimators are denoted as $\bF_{1}, \ldots, \bF_{k}$ and $\bLambda_{1}, \ldots, \bLambda_{k}$.

\item[{\bf Step 5.}] Given $\bLambda_{1}, \ldots, \bLambda_{k}, \bF_{1}, \ldots, \bF_{k}$ and $g_{i},\;i=1\ldots,N$, we can define
$$\by_{i}-\bF_{g_{i}}^{'}\blambda_{g_{i},i} = \bx_{i}^{'}\bbeta+\bvarepsilon_{i},\;i=1,\ldots,N,$$
then the updated least squares estimator $\bbeta^{(h+1)}$ can be obtained and set $h = h + 1$.

\item[{\bf Step 6.}] Repeat Step 2 -- Step 5 until convergence occurs.
\end{itemize}

Although the least squares objective function is not globally convex \citep{Bai2009}, from the results of the Monte Carlo simulation studies, the proposed algorithm is robust to the starting value under large $N$ and large $T$. In the numerical experiments, we propose using the least squares method to get the initial value $\bbeta^{(0)}$ by ignoring the unobserved group factor structures. We observe in the Monte Carlo simulation studies (Section 5) that the initial value $\bbeta^{(0)}$ has good convergence property. In practice, if one has concern that the algorithm may converge to a local optimizer, we suggest using different random starting values, and then select the solution that yields the lowest value of the objective function if those solutions are different.

%The complexity of the above iterative procedure is $O(TNr)+O(\delta_{NT}^{3})$, where $\delta_{NT}=\min[N,T]$.
To evaluate the complexity of the above iterative procedure, we consider the complexity of Steps 2--4 in the above algorithm. In Step 2, we need to obtain the top $r$ singular values of the $T \times N$ data matrix in order to obtain the values $\bF$ and $\bLambda$, which requires a complexity of $O(\delta_{NT}^2\tilde{\delta}_{NT})+O(\delta_{NT}^3)$ where $\delta_{NT}=\min[N,T],\tilde{\delta}_{NT}=\max[N,T]$. Step 3 of the algorithm requires sorting the $\blambda_i$ according to their projection distances to these subspaces; the computational cost is $O(Nkr)$.
In Step 4, the least squares estimation requires a complexity of $O(p^2N)$ because of the inverse operation. Thus, the complexity of the above iterative process is $O(\delta_{NT}^2\tilde{\delta}_{NT})+O(\delta_{NT}^3)+O(Nkr)+O(p^2N)=O(\delta_{NT}^2\tilde{\delta}_{NT})$.

\subsection{Subspace Clustering for Factor Loadings}

In this section, we present the procedure of the subspace segmenting {for the factor loadings $\blambda_{i},\;i=1,\ldots,N$. We assume that the $i$-th factor loading $\blambda_{i}\in \mathbb{R}^{r}$ ($i=1,\ldots,N$) is inside $k$ different subspaces, in which the dimension of these $k$ subspaces are $d_{1}, d_{2}, \ldots, d_{k}$, where $\;0 < d_{j} < r, \;j=1,2, \ldots, k$. For the $j$-th subspace $S_{j} \subset \mathbb{R}^{r}$ with dimension $d_{j}$, a basis $B_{j} = [\bb_{j1},\ldots,\bb_{j,r-d_{j}}]\in \mathbb{R}^{r\times (r-d_{j})}$ is selected for its orthogonal complement $S_{j}^{\perp}$.} Using these notations, we can obtain the following equation for the $j$-th subspace $S_{j}$ and the $i$-th factor loading $\blambda_i$:
\begin{equation}
 \{\blambda_i\in S_{j}\} =\{\blambda_i\in \mathbb{R}^{r}:B_{j}^{T}\blambda_i=0  \} = \left\{\blambda_i\in \mathbb{R}^{r}:\bigwedge_{m=1}^{r-d_{j}}(\bb_{jm}^{T}\blambda_i=0)  \right\}.
\label{3.4}
\end{equation}
Since $\blambda_i \in \mathbb{R}^{r}$ belongs to $\cup_{j=1}^{k}S_{j}$ if and only if $( \blambda_i \in S_{1}) \vee \ldots \vee (\blambda_i\in S_{k})$, where the notation $\bigvee$ represents the ``or" operator.
This condition is equivalent to
\begin{equation}
\bigvee_{j=1}^{k}(\blambda_i\in S_{j})\Leftrightarrow \bigvee_{j=1}^{k}\bigwedge_{m=1}^{r-d_{j}}(\bb_{jm}^{T}\blambda_i=0)\Leftrightarrow \bigwedge_{\sigma}\bigvee_{j=1}^{k}(\bb_{j\sigma(j)}^{T}\blambda_i = 0),
\label{3.5}
\end{equation}
where the notation $\bigwedge$ represents the ``and" operator, and $\sigma$ is a particular choice of the  normal vector $\bb_{j\sigma(j)}$ from the $j$-th basis $B_{j}$. Note that the right-hand side of Eq. (\ref{3.5}) is obtained by exchanging the ``and" and ``or" operators using De Morgan's laws. Since
\begin{equation}
\bigvee_{j=1}^{k}(\bb_{j\sigma(j)}^{T}\blambda_i=0) \Leftrightarrow \prod_{j=1}^{k} (\bb_{j\sigma(j)}^{T}\blambda_i=0) \Leftrightarrow p_{k\sigma}(\blambda_i)=0,
\label{3.6}
\end{equation}
 which is a homogeneous polynomial of degree $k$ in $r$ variables, we can write each of the polynomials as
\begin{equation}
p_{k\sigma}(\blambda_i)=\prod_{j=1}^{k}(\bb_{j\sigma(j)}^{T}\blambda_i)=\bc_{k}^{T}\bv_{k}(\blambda_i)=
\sum\limits_{\stackrel{0 \leq k_{s} \leq k, s = 1, 2, \ldots, r}{\sum_{s=1}^{r} k_{s} = k}}
c_{k_{1},k_{2},\ldots,k_{r}}\lambda_{i1}^{k_{1}}\lambda_{i2}^{k_{2}}\ldots \lambda_{ir}^{k_{r}}=0,\
\label{3.7}
\end{equation}
 where $\bc_{k}$ is the vector of polynomial coefficients,
$c_{k_{1},k_{2},\ldots,k_{r}}$ is the polynomial coefficient,
%and $0 \leq k_{s} \leq k\;,s = 1, 2, \ldots, r$ and $\sum_{s=1}^{r}k_{s} = k$,
$\blambda_i = (\lambda_{i1},\cdots,\lambda_{ir})$, $\bv_{k}: \mathbb{R}^{r}\rightarrow \mathbb{R}^{M_{k}(r)}$ is the Veronese map of degree $k$ \citep{Fischler1981} which is also known as the polynomial embedding in machine learning defined as $\bv_{k}: [\lambda_{i1}, \ldots, \lambda_{ir}]^{T} \mapsto [\ldots,\blambda_i^{I},\ldots]^{T}$ with $I$ being chosen in the degree-lexicographic order, $\blambda_i^{I}=\lambda_{i1}^{k_{1}}\lambda_{i2}^{k_{2}}\ldots \lambda_{ir}^{k_{r}}$ and the dimension $M_{k}(r)=C_{k+r-1}^{r-1}$.

Since the polynomial in Eq. (\ref{3.7}) can be satisfied by all the factor loadings $\blambda_{i}, i = 1, 2, \ldots, N$, we can then use these factor loadings to obtain the subspaces. Although the polynomial equations in Eq. (\ref{3.7}) are nonlinear in each point $\blambda_i$, these polynomials are actually linear in the vector of polynomial coefficients $\bc_{k}$.
Indeed, since each polynomial $p_{k \sigma}(\blambda_i)=\bc_{k}^{T}\bv_{k}(\blambda_i)$ must be satisfied by every data point, we can obtain $\bc_{k}^{T}\bv_{k}(\blambda_{i})=0$ for all $i = 1, 2, \ldots, N$.

Suppose that $I_{k}$ is the space of the vector of polynomial coefficients $\bc_{k}$ of all the homogeneous polynomial that vanishes in the $k$ subspaces, then the vector of polynomial coefficients of the factorizable polynomial defined in Eq. (\ref{3.6}) span a (possibly proper) subspace in $I_{k}$ as
$span_{\sigma}\{ p_{k\sigma} \}\subseteq I_{k}$. As every vector $\bc_{k}$ in $I_{k}$ represents a polynomial that vanishes on all the data points (on the subspaces), the vector $\bc_{k}$  must satisfy the system of linear equations
\begin{equation}
\bc_{k}^{T}V_{k}(r)=\bc_{k}^{T}[\bv_{k}(\blambda_{1}),\ldots,\bv_{k}(\blambda_{N})]={\bf 0}^{T},
\label{3.10}
\end{equation}
where $V_{k}(r)\in \mathbb{R}^{M_{k}(r)\times N}$ is the embedded data matrix.Hance, we have the relationship $I_{k} \subseteq null(V_{k}(r))$.

\noindent
\textbf{Remark 1:} The zero set of each vanishing polynomial $p_{k}(\blambda_i),\;i=1, 2, \ldots, N$ is a surface in $\mathbb{R}^{r}$, therefore, the derivative of $p_{k}(\blambda_i)$ at $\blambda_{i} \in S_{j}$, denoted as $Dp_{k}(\blambda_i)$, gives a vector normal to the surface. Since a union of subspaces is locally flat, i.e., in a neighborhood of $\blambda_{i}$ the surface is merely the surface $S_{j}$, then the derivative at $\blambda_{i}$ lies in the orthogonal complement $S_{j}^{\perp}$ of $S_{j}$. By evaluating the derivatives of all the polynomials in $I_{k}$ at the same point $\blambda_{i}$, we obtain a set of normal vectors that span the orthogonal complement of $S_{j}$.
%Thus, we have the following theorem \cite{Vidal}:
%\noindent
%\begin{theorem}
% (Obtaining subspace by polynomial differentiation) Let $I_{k}$ be (the space of coefficient vectors of) the set of polynomials of degree $k$ that vanish on the $k$ subspaces. If the data set $\bX$ satisfies $dim(null(V_{k}(r))) = dim(I_{k})$ and one generic point $\by_{i}$ is given for each subspace $S_{j}$, then we have
%\begin{equation}
%S_{j\perp} = span \left\{ \left. \frac{\partial c_{k}^{'}v_{k}(\bx)}{\partial \bx} \right|_{\bx=\by_{i}},\forall c_{k}\in null(V_{k}(r)) \right\},
%\label{3.12}
%\end{equation}
%where $S_{j\bot}$ represents the normal vectors of the subspace $S_{j}$.
%\end{theorem}

Following the results in Theorem 3 of \cite{Vidal2005}, we can obtain a set of polynomial $p_k(\blambda_i),i=1\cdots,N$ with coefficients equal to the eigenvectors in the null space of $V_{k}(r)$.
 By evaluating the derivatives $Dp_k(\blambda_i)$ at each $\blambda_i$, $i = 1, \cdots,N$, we can obtain a set of vectors orthogonal to the subspace that the points lie in.
Note that the generalized principal component analysis method \citep{Vidal2005,Vidal2016} relies on reliable samples per subspace to segment the dataset, however, in the presence of noise, the sample may not be reliable. %Here, we consider the factor loadings as samples.
%But if we take the advantage of the normals of all samples in one subspace, we expect that it will smooth out the random noise. More specifically,
Here, for each sample, we assume that the sample could be obtained from all the candidate co-dimension classes, and the sample is voted by the dominate vectors of $Dp_k(\blambda_i)$ as a basis.
Finally, the base associated with the highest vote will be used as the normal vectors perpendicular to the subspaces as suggested by \cite{Yang2005}.
After obtaining the orthogonal bases of those subspaces, we can assign $\blambda_{i}$ to the subspace $j^{*}$, where
$j^{*} = \arg \min_{j =1,\ldots,k}|| B_{j}^{T}\blambda_{i}||$.

%\textbf{Remark 2:} The common factor $\bF$ is obtained by the principal components analysis from the matrix $\bY^{*}\bY^{*'}/N$. When $N$ and $T$ are large, $\bY^{*}\bY^{*'}/N \rightarrow \bF\Sigma_{\Lambda}\bF^{'}+\Omega$, where $\Sigma_{\Lambda}$ is the limit of the $r\times r$ matrix $\bLambda^{'}\bLambda/N$ and $\Omega$ is a $T \times T$ diagonal matrix with equal elements of the covariance matrix of $\bvarepsilon_{i}$.

\section{Asymptotic Properties}
In this section, we characterize the asymptotic properties of the estimators as $N$ and $T$ tend to infinity. We prove that the estimated clustering converges to the corresponding true subspaces under some conditions.
In particular, we follow the method in \cite{Pollard1981} to establish the consistency of subspace clustering.

\subsection{Consistency of clustering procedure}
The proposed clustering procedure prescribes a criterion for partitioning a set of points into $k$ subspaces. To divide the factor loadings $\blambda_{1}, \ldots, \blambda_{N}$ in $\mathbb{R}^{r}$, we first choose $k$ ($k$ is fixed) cluster subspaces $S_{1},S_{2},\ldots,S_{k}$ with dimensions $d_{1}, d_{2}, \ldots, d_{k}$, respectively, that minimize
\begin{equation}
W_{N} = \frac{1}{N}\sum_{i=1}^{N}\min_{1\leq j\leq k}\phi(\Delta( \blambda_{i},S_{j} )),
\label{5.1}
\end{equation}
where $\blambda_{1}, \ldots, \blambda_{N}$ can be viewed as $N$ vectors of the sample points and
$\Delta( \blambda_{i},S_{j}  )$ is the angle between the vector $\blambda_{i}$ and the subspace $S_{j}$ which is a value in $[0, \pi/2]$.
Let $|| \blambda_{i} ||=1$ and $\bb_{j1},\ldots,\bb_{jm}$ be an orthonormal basis of $S_{j}$ and $\theta$ be the angle between $\blambda_{i}$ and $S_{j}$, then $\sin(\theta)=|| \blambda_{i}-\sum_{\ell=1}^{m}\bb_{j \ell}(\bb_{j \ell}^{T}\blambda_{i}) ||=\sqrt{1-\sum_{\ell=1}^{m}(\bb_{ j\ell}^{T}\blambda_{i})^{2}}$. Thus, clustering by angles is equivalent to clustering by projecting the $\blambda_{i}$ to the nearest subspace.
Here, the function $\phi$ must satisfy some regularity conditions that $\phi$ is continuous and nondecreasing, with $\phi(0) = 0$. For any subspaces $\bx$ and $\by$, $\Delta(\bx, \by) \in [0,\pi/2]$, therefore, $\phi(\Delta( \bx,\by ))$ must be in a compact set.

Since $\max\limits_{1\leq i \leq N}||\hat{\blambda}_i-\blambda_i||=o_{p}(1)$, we can show that the empirical distribution function converge uniformly to the true distribution function by the strong law of large numbers and the Glivenko-Cantelli theorem, i.e.,
$$\lim\limits_{N\rightarrow \infty} \sup\limits_{\bx\in \mathbb{R}^{r}}|\hat{P}_{N}(\bx)-P(\bx)|=0,$$
where $\hat{P}_{N}(\bx)=\frac{1}{N}\sum\limits_{i=1}^{N}I_{\{ \hat{\blambda}_{i}\leq \bx \}}$, with $I_{A} = 1$ if $A$ is true and 0 otherwise, is the empirical distribution function and $P(\bx)$ is the true distribution function. Therefore, clustering for the estimate $\hat{\blambda}_{i}$ obtained by our proposed method is equivalent to clustering for the true value $\blambda_{i}$ ($i = 1, 2, \ldots, N$).
Since $\phi(\Delta( \blambda,S ))$ is an increasing function of the angle deviation which can be used in defining a within cluster sum of angle deviations, the criterion considered here minimizes the within-cluster sum of angle deviations.

We assume that $\{\blambda_{1},\ldots,\blambda_{N}\}$ is a sample of independent observations on some probability measure $P$. Here, we consider the empirical measure
\begin{equation}
W(\mathcal{S}, {\hat P}_{N})=\int \min_{S\in \mathcal{S}}\phi(\Delta( \blambda_{i},S )) {\hat P}_{N} (d\blambda),
\label{5.2}
\end{equation}
where $\mathcal{S}$ is a set of subspaces. For a fixed set of subspaces $\mathcal{S}$, we can obtain
\begin{equation}
W(\mathcal{S}, {\hat P}_{N}) \xrightarrow{a.s.} W(\mathcal{S},P)=\int \min_{S\in \mathcal{S}}\phi(\Delta( \blambda_{i},S ))P (d\blambda),
\label{5.3}
\end{equation}
where $\blambda_{1},\ldots,\blambda_{N}$ are the $N$ vectors of the sample points which can be clustered by minimizing the within cluster sum of angle deviations.
Let $\mathcal{S}_{N}$ be the set of subspaces that minimizes $W(\cdot, {\hat P}_{N})$ (i.e., the set of optimal clustered subspaces based on the samples) and $\bar{\mathcal{S}}$ be the set of subspaces that minimizes $W(\cdot, P)$. Provided that  $\bar{\mathcal{S}}$ can be uniquely determined, it is expected that $\mathcal{S}_{N}$ should lie close to $\bar{\mathcal{S}}$.

For a probability measure $Q$ on $\mathbb{R}^{r}$ and a finite set of subspaces $\mathcal{S}$ of $\mathbb{R}^{r}$, we define
\begin{equation}
\Phi(\mathcal{S},Q)=\int \min_{S\in \mathcal{S}}\phi (\Delta( \blambda,S  ))Q(\blambda)
\label{5.4}
\end{equation}
and
\begin{eqnarray}
%\nonumber
m_{k}(Q) =  \inf \{\Phi(\mathcal{S},Q): \text{$\mathcal{S}$ contains $k$ or fewer subspaces};\text{and $d_{1},  \ldots, d_{k}$ are  known}\}.
\label{5.5}
\end{eqnarray}
For a given value of $k$, the set of optimal clustered subspaces based on the samples $\mathcal{S}_{N} = \mathcal{S}_{N}(k)$ is chosen to satisfy $\Phi(\mathcal{S}_{N}, {\hat P}_{N})=m_{k}({\hat P}_{N})$ and the set of optimal population clustered subspaces $\bar{\mathcal{S}}=\bar{\mathcal{S}}(k)$ is chosen to satisfy $\Phi(\bar{\mathcal{S}},P)=m_{k}(P)$.

To define the distance measures, we have the following assumption:\\
\noindent
\textbf{Assumption A.} Suppose that $\int \phi(|| x ||)P(dx)<\infty$ and that $m_j(P)>m_k(P)$ for $j=1,\cdots,k-1$.

Our aim here is to prove a consistency result for the cluster subspaces that $\mathcal{S}_{N} \xrightarrow{a.s.} \bar{\mathcal{S}}$. To show $\mathcal{S}_{N}\xrightarrow{a.s.}\bar{\mathcal{S}}$, we first consider the subspace distance defined in \cite{LiWang2006}:

\noindent
\textbf{Definition 1.} The symmetric distance between any $m$ dimensional subspace $U$ and $n$-dimensional subspace $\tilde{U}$ is defined as
\begin{eqnarray}
\nonumber
D(U, \tilde{U}) & = & \max(\vec{D}(U,\tilde{U}),\vec{D}(\tilde{U},U)) %= \sqrt{\sum_{i=1}^{m}D^{2}(\bu_{i}, \tilde{U})}
= \sqrt{\max(m,n)-\sum_{i=1}^{m}\sum_{j=1}^{n}(\tilde{\bu}_{j}^{'}\bu_{i})^{2}},
\label{5.6}
\end{eqnarray}
where $(\bu_{1},\ldots,\bu_{m})$ and $(\tilde{\bu}_{1},\ldots,\tilde{\bu}_{n})$ are the bases of $U$ and $\tilde{U}$ respectively. Note that this subspace distance satisfies the triangle inequality
\begin{eqnarray*}
D(U,\tilde{U}) \leq D(U,W) + D(W,\tilde{U}),
\end{eqnarray*}
where $W$ is any non-empty subspace.

\noindent
\textbf{Remark 2:} The angle $\Delta(\cdot, \cdot)$ used in Eq. (\ref{5.1}), $$\Delta(U,\tilde{U})=\sqrt{\min(m,n)-\sum_{i=1}^{m}\sum_{j=1}^{n}(\tilde{\bu}_{j}^{'}\bu_{i})^{2}},$$ is different from the distance defined in Definition 1 for measuring the subspaces distance.
In fact, the angle $\Delta(\cdot, \cdot)$ projects the low subspace onto the high-dimensional subspace, while the distance $D(\cdot,\cdot)$ projects the high-dimensional subspace onto the low subspace.
%Thus, the difference between Min and Max will occur.
In order to avoid different dimension subspaces being treated as the same subspace, $D(\cdot, \cdot)$ is used to characterize the difference between two subspaces.
For example, consider a two-dimensional plane and a line which is parallel to this plane as two subspaces, if we use the distance $\Delta(\cdot, \cdot)$ to characterize the difference between these two subspaces, it is likely to get the result that these two subspaces are treated as the same subspace because the angel between these two subspaces is $0$. However, using the distance $D(\cdot, \cdot)$ can avoid this issue. Here, we further define a distance measure similar to the Hasudroff distance:\\
\noindent
\textbf{Definition 2.} Let $\mathcal{X}$ and $\mathcal{Y}$ be two non-empty compact subsets that contain multiple subspaces. We define their Hausdorff distance $D_{H}(\mathcal{X}, \mathcal{Y})$ by
\begin{equation}
D_{H}(\mathcal{X}, \mathcal{Y}) = \max \{ \sup_{x\in \mathcal{X}}\inf_{y\in \mathcal{Y}}D(x,y), \sup_{y\in \mathcal{Y}}\inf_{x\in \mathcal{X}}D(x,y)\},
\label{5.7}
\end{equation}
where $x \in \mathcal{X}$ is a subspace rather than a point, and the $D(\cdot,\cdot)$ is the distance between two subspaces defined in Definition 1.

By Definition 2, we have $D_{H}(\mathcal{X},\mathcal{Y}) < \delta$ if and only if every subspace of $\mathcal{X}$ is within the distance $\delta$ of at least one of the subspaces of $\mathcal{Y}$, and vice versa. Suppose $\mathcal{X}$ contains exactly $k$ distinct subspaces, and that $\delta$ is chosen to be a value less than half of the minimum distance between the subspaces of $\mathcal{X}$. Then, if $\mathcal{Y}$ is any set of $k$ or fewer subspaces for which $D_{H}(\mathcal{X},\mathcal{Y})< \delta$, the $\mathcal{Y}$ must contain exactly $k$ distinct subspaces. Therefore, the almost sure convergence of $\mathcal{X}_{N}$ in the above sense of distance could be translated into the almost sure convergence of subspaces with a suitable labeling. By definition, for any two subspaces $S_{1}$ and $S_{2}$, if $D(S_{1},S_{2}) < \delta$, then $\Delta( S_{1},S_{2})<\delta$. We will provide the following theorem for the consistency of clustering procedure. Since the conclusion of the theorem is in terms of almost sure convergence, there might be aberrant null sets of subspaces $S$'s for which the convergence does not hold. In order to estimate the parameters in the model presented in Eq. (\ref{2.1}) and to prove the consistency of the estimators, similar to \cite{Bonhomme2015} and \cite{Ando2016}, we add the following assumption, Assumption B, that each group must have a certain proportion individuals. This assumption also guarantees the uniqueness because the null set situation is excluded.

\noindent
\textbf{Assumption B.} All units are divided into a finite number of subspaces $k$, each of them containing $N_j$ units such that $0<\underline{a}<N_j/N<\bar{a}<1$.

For notational simplicity and clarity, we assume that $\lambda_i, i = 1, \cdots, N$ is known in the following theorem.
\begin{theorem}{Theorem 1.}{}
Suppose Assumptions A and B hold and for each $j = 1, 2, \ldots, k$ there exists an unique set of subspaces $\bar{\mathcal{S}}(j)$ that satisfies $\Phi(\bar{\mathcal{S}}(j),P)=m_{j}(P)$, then,
$\mathcal{S}_{N} \xrightarrow{a.s.} \bar{\mathcal{S}}(k)$, and $\Phi(\mathcal{S}_{N},P_{N})\xrightarrow{a.s.}m_{k}(P)$.
\end{theorem}
The proof of Theorem 1 is presented in the Supplementary Materials S2.

\subsection{Consistency of the estimators}
In this subsection, we discuss the asymptotic properties of the proposed estimators. Recall that the proposed estimators can be obtained by minimizing the objective function defined in Eq. (\ref{objfct}) subject to the constraints $g_{i}=\arg\min\limits_{j \in\{1,\ldots,k\}} || B_{j}^{T}\blambda_{i}||$, $\bF_{j}^{'}\bF_{j}/T=I_{r}(j=1,\ldots,k),\bLambda_{j}^{'}\bLambda_{j}(j=1,\ldots,k)$ being diagonal.
%This optimization problem needs to use the least squares estimation method and subspace clustering simultaneously.
While the consistency of the subspace clustering procedure has been discussed in Section 4.1, we have the following theorems (Theorems 2, 3 and 4) to show the property of the estimators when $T$ and $N$ are large.
To show the property of the estimators, Assumptions C, D, E and F, are needed as presented in \cite{Bai2009} and \cite{Ando2016} and we present Assumptions C--F in the Supplementary Materials S1.

\begin{theorem}{Theorem 2.}{} Suppose that Assumptions A--E hold, as $N \rightarrow \infty$ and $T \rightarrow \infty$, the following statements hold:
\begin{itemize}
\item[(i)] $||\hat{\bbeta}-\bbeta^{0}||=o_{p}(1)$,
\item[(ii)] $||P_{\hat{\bF}_{j}}-P_{\bF^{0}_{j}}||=o_{p}(1),j=1,\ldots,k.$
\end{itemize}
\end{theorem}

\begin{theorem}{Theorem 3.}{} Consistency of the estimator of group membership: Suppose that Assumptions A--E hold, then for all $\tau > 0$ and $T, N \rightarrow \infty$, we have
\begin{eqnarray*}
P\left(\sup\limits_{i\in\{1,\cdots,N\}} \left| \hat{g}_i(\hat{\bbeta},\hat{\bF},\hat{\bLambda},\hat{B}_1,\cdots,\hat{B}_k)-g_i^0 \right| \right)=o(1)+o(N/T^{\tau}).
\end{eqnarray*}
\label{group}
\end{theorem}

\begin{theorem}{Theorem 4.}{} Asymptotic normality: Suppose that Assumptions A--F hold and $T/N \rightarrow \rho>0$, then
\begin{eqnarray*}
\sqrt{NT}(\hat{\bbeta}-\bbeta^0)\rightarrow^d N(v_0,V_\beta(\bF_1^0,\cdots,\bF_k^0,B^0_1,\cdots,B^0_k)),
\end{eqnarray*}
where $v_0$ is the probability limit of
\begin{eqnarray*}
v & = & \sqrt{\frac{T}{N}}\sum\limits_{j=1}^{k}D(F_1^0,\cdots,F_k^0,B_1^0,\cdots,B_k^0)^{-1}\eta_j +\sqrt{\frac{T}{N}}\sum\limits_{j=1}^{k}D(F_1^0,\cdots,F_k^0,B_1^0,\cdots,B_k^0)^{-1}\zeta_j
\end{eqnarray*}
with
\begin{eqnarray*}
\eta_j&=& -\frac{1}{N_j}\sum\limits_{i:g_i^0=j}\sum\limits_{\ell:g_\ell^0=j}
\frac{(x_i-V_{j,i})^{'}F_j^0}{T} \left(\frac{F^{0'}_{j}F^{0}_{j}}{T} \right)^{-1}\left(\frac{\Lambda_j^{0'}\Lambda_j^{0}}{N_j} \right)^{-1}
\lambda_{g_\ell^0,\ell}\left(\frac{\varepsilon_i^{'}\varepsilon_\ell}{T}\right),\\
\zeta_j&=& \sum\limits_{j=1}^{k}\frac{1}{N_j T}\sum\limits_{i:g_i^0=j}\sum\limits_{\ell:g_\ell^0=j}x_i^{'}M^0_j \Omega_kF^0_j(F^{0'}_{j}F^0_j/T)^{-1} (\Lambda_j^{0'}\Lambda_j^{0}/N_j)^{-1}\lambda^0_{j,i},
\end{eqnarray*}
\begin{eqnarray*}
 D(F_1^0,\cdots,F_k^0; B_1^0,\cdots,B_k^0)& = & \frac{1}{NT}\sum\limits_{j=1}^{k}\sum\limits_{i:g_i^0=j}x_i^{'}M^0_jx_i-\frac{1}{NT}\sum\limits_{j=1}^{k} \left[\frac{1}{N_j}\sum\limits_{i:g_i^0=j}\sum\limits_{\ell:g_\ell^0=j}x_i^{'}M^0_jx_\ell c_{j,\ell i} \right],
\end{eqnarray*}
\begin{eqnarray*}
 V_\beta(F_1^0,\cdots,F_k^0;B_1^0,\cdots,B_k^0)&=&D_0(F_1^0,\cdots,F_k^0;B_1^0,\cdots,B_k^0)^{-1}J_0(F_1^0,\cdots,F_k^0;B_1^0,\cdots,B_k^0)\\
&&D_0(F_1^0,\cdots,F_k^0;B_1^0,\cdots,B_k^0)^{-1},
\end{eqnarray*}%}
where $D_0(F_1^0,\cdots,F_k^0;B_1^0,\cdots,B_k^0)$ is the probability limit of $D(F_1^0,\cdots,F_k^0;B_1^0,\cdots,B_k^0)$ and $J_0(F_1^0,\cdots,F_k^0;B_1^0,$ $\cdots,B_k^0)$ is defined in Assumption F and $V_{j,i}=N_j^{-1}$ $\sum\limits_{\ell:g_\ell^0=j}c_{j,\ell i}x_\ell x_i$, and $M^0_j=\frac{1}{T}F^0_{j}B^0_jB_j^{0T}F^{0T}_{j}$.
\label{normality}
\end{theorem}
The proofs of Theorems 2--4 are presented in the Supplementary Materials S3.

\section{Monte Carlo Simulation Studies}

In this section, Monte Carlo simulation studies with different settings are used to illustrate the proposed methodologies and to study the finite sample properties of the proposed methods. We assume that $\blambda_{i}$ ($i = 1, 2, \ldots, N$) comes from different subspaces with dimension $d_{1}, d_{2}, \ldots d_{k}$ and $\blambda_{i}$ can follow different probability distributions over different subspaces. Furthermore, we consider that $\blambda_{i}, i=1,2,\ldots, N$, can have moderate noise. The simulation results are based on $100$ simulations.

\subsection{Setting 1}
We consider the situation that there are three different subspaces in the $\mathbb{R}^{3}$ space with known dimensions $d_{1}$, $d_{2}$ and $d_{3}$, i.e., $k = 3$ and $r = 3$. The bases of the three subspaces with dimensions $d_{1}$, $d_{2}$ and $d_{3}$ are represented as $\balpha_{1}$, $\balpha_{2}$ and $\balpha_{3}$, respectively. Let $N_{\ell_1 \times \ell_2}(\mu, \sigma^{2})$ be a $\ell_1 \times \ell_2$ matrix whose elements are random variables that are independent and identically distributed as normal with mean $\mu$ and variance $\sigma^{2}$.
Under this setting, we generate the panel data $y_{it}$ in the $j$-th subspace ($j = 1, 2, 3$; $i = 1, 2, \ldots, N_{j}$; $t = 1, 2, \ldots, T$) based on the panel data model in Eq. (\ref{2.1}) with $N_{1} = N_{2} = N_{3} = N = 100$, $T = 6$ and the following scheme:
\begin{itemize}
\item $\blambda_{i} \sim N_{r \times 1}(1, 1)$ with random noise from a normal distribution with mean 0 and variance 0.1, i.e.,
$\bLambda_{j} = N_{N_{j} \times d_{j}}(1, 1) \balpha_{j}' + N_{N_{j} \times r}(0, 0.1),\balpha_{j}~\sim N_{r \times d_j}(0, 1)~j = 1, 2, 3;$
\item $p = 2$ with $\bbeta = (\beta_{1}, \beta_{2})' = (1, 2)'$;
\item the covariate $\bX$ is a $T\times N \times p$ array with
\begin{eqnarray*}
\bX_{\cdot \cdot 1}  =  \bmu_{1} + c_{1}\bF \bLambda' + \btau \bLambda' + \beeta_{1}
{\mbox {, }} \bX_{\cdot \cdot 2}  =  \bmu_{2} + c_{2}\bF \bLambda' + \btau \bLambda' + \beeta_{2},
\end{eqnarray*}
where $\bX_{\cdot \cdot 1}$ and $\bX_{\cdot \cdot 2}$ are $T \times N$ matrices, $\bmu_{1}, \bmu_{2}$ are $T \times N$ matrices of all the elements that are $1$, and $c_{1}=c_{2}=0.5$;
\item $\bF\bLambda = (\bF_{1} \bLambda_{1}^{T}, \bF_{2} \bLambda_{2}^{T},\bF_{3}\bLambda_{3}^{T})$,
where $\bF_{1},\bF_{2},\bF_{3}$ are $T \times r$ matrices that satisfy $\bF_{j}^{'}\bF_{j}/T=I_{r},\;j=1,2,3$;
\item $\beeta_{1} \sim N_{T \times N}(0, 1)$ and $\beeta_{2} \sim N_{T \times N}(0, 1)$;
\item $\btau$ is a $T\times r$ matrix of all the elements that are $1$;
\item the random error $\varepsilon_{i,t} \stackrel{\text{i.i.d}}{\sim} N(0, 0.5)$,~$i = 1, 2, \ldots, N, t = 1, 2, \ldots, T$.
\end{itemize}

In the simulation study, we compare the performance of the proposed LSSC method with the GFE method  \citep{Bonhomme2015} and the estimation method proposed by \cite{Bai2009} (BAI) in terms of the biases and the root mean squared errors (RMSEs). The simulated biases and RMSEs of the estimators obtained from the GFE, BAI and LSSC estimation methods for Setting 1 are presented in Table \ref{example1}. From Table \ref{example1}, we observe that the proposed LSSC method has better performance than the GFE and BAI's methods in terms of biases and RMSEs. It is noteworthy that although the theoretical proofs of the asymptotic properties require $T$ to be large, the simulation results show that the proposed method performs well even when $T$ is small.

\begin{table*}
\caption{Simulated biases and root mean squared errors (RMSEs) of the GFE, BAI and LSSC estimation methods for Setting 1}
\centering
\label{example1}
\begin{center}
\begin{tabular}{l l r r r r r r} \hline
         &  & \multicolumn{2}{c}{GFE} &  \multicolumn{2}{c}{BAI}  & \multicolumn{2}{c}{LSSC} \\
Dimension of subspaces  &  &  Bias & RMSE & Bias & RMSE &  Bias & RMSE \\ \hline
$d_{1} = d_{2} = d_{3} = 1$ & $\beta_{1}$ & 0.1271 & 0.1304 & 0.0067 & 0.0121 & 0.0010 & 0.0035\\
                            & $\beta_{2}$ & 0.1255 & 0.1295 & 0.0066 & 0.0116 & 0.0011 & 0.0035\\ \hline
$d_{1} = d_{2} = d_{3} = 2$ & $\beta_{1}$ & 0.2042 & 0.2082 & 0.2821 & 0.3141 & 0.0225 & 0.0518\\
                            & $\beta_{2}$ & 0.2042 & 0.2084 & 0.2819 & 0.3120 & 0.0218 & 0.0503\\ \hline
$d_{1} = d_{2} = 2, d_{3} = 1$ & $\beta_{1}$ & 0.1530 & 0.1580 & 0.1871 & 0.2258 & 0.0084 & 0.0765\\
 & $\beta_{2}$ & 0.1571 & 0.1620 & 0.1884 & 0.2269 & 0.0111 & 0.0740\\ \hline
$d_{1} = 2, d_{2} = d_{3} = 1$ & $\beta_{1}$ & 0.1770 & 0.1793 & 0.0898 & 0.1272 & 0.0034 & 0.0144\\
 & $\beta_{2}$ & 0.1786 & 0.1810 & 0.0906 & 0.1262 & 0.0036 & 0.0149\\ \hline
\end{tabular}
\end{center}
\end{table*}
To compare the performance of the clustering methods, we present the simulated average misclassification rates of the clustering methods based on GFE and the proposed LSSC in Table \ref{example1_2}. From Table \ref{example1_2}, we can see that the simulated misclassified rates of the LSSC method are lower than the corresponding misclassified rates of the GFE method.
\begin{table*}
\centering
\caption{Simulated average misclassified rate of the GFE and LSSC clustering methods for Setting 1}
  \begin{center}
%Compared our proposed methods with the randomized exchange algorithm (REX) for $D$-optimality.
\begin{tabular}{lll} \hline
Dimension of subspaces    &  GFE  &   LSSC    \\  \hline
$d_{1} = d_{2} = d_{3} = 1$ &0.2982 &  0.0660  \\
$d_{1} = d_{2} = d_{3} = 2$ &0.2713 &  0.0908 \\
$d_{1} = d_{2} = 2, d_{3} = 1$&0.3428 &  0.1511\\
$d_{1} = 2, d_{2} = d_{3} = 1$ &0.3314 &  0.1416\\ \hline
\end{tabular}
\end{center}
\label{example1_2}
\end{table*}
To verify the consistency of the proposed LSSC method, in addition to $N_{1} = N_{2} = N_{3} = N = 100$, we consider different sample sizes $N = 50$, 200, 300 and 500 in order to study the effect of the sample size on the biases and RMSEs. We also consider different values of time period $T = 5$, 10, 30, 50, 100 to study the effect of the time period on the biases and RMSEs of parameter $\bbeta$. The results are presented in Tables \ref{example1_3} and \ref{example1_4}. From Tables \ref{example1_3} and \ref{example1_4}, we can see that the biases and RMSEs of the LSSC estimators decrease as the sample size $N$ or time period $T$ increases, which verifies Theorem 2.
\begin{table*}
\centering
\caption{Simulated biases and root mean squared errors (RMSEs) of the LSSC estimator with $d_{1}=2,d_{2}=2,d_{3}=2$ in Setting 1 with different sample sizes $N_{1} = N_{2} = N_{3} = N$}
\begin{center}
\begin{tabular}{lrrrr} \hline
              & \multicolumn{2}{c}{$\beta_{1}$}  & \multicolumn{2}{c}{$\beta_{2}$} \\
Sample size ($N$)    &  Bias & RMSE &  Bias & RMSE \\ \hline
%$n_{1}=n_{2}=n_{3}=30$ & 0.0064 & 0.0061 & 0.0006 & 0.0006\\
$50$  & 0.0701 & 0.0691 & 0.0304 & 0.0304\\
$100$ & 0.0525 & 0.0524 & 0.0225 & 0.0227\\
$200$ & 0.0525 & 0.0537 & 0.0214 & 0.0212\\
$300$ & 0.0378 & 0.0381 & 0.0167 & 0.0169\\
$500$ & 0.0018 & 0.0021 & 0.0008 & 0.0011\\ \hline
\end{tabular}
\end{center}
\label{example1_3}
\end{table*}

\begin{table*}
\centering
\caption{Simulated biases and root mean squared errors (RMSEs) of the LSSC estimator with $d_{1}=2,d_{2}=2,d_{3}=2$ in Setting 1 with different time period $T$ and same sample sizes $N_{1} = N_{2} = N_{3} = N$}
\begin{center}
\begin{tabular}{lrrrr} \hline
              & \multicolumn{2}{c}{$\beta_{1}$}  & \multicolumn{2}{c}{$\beta_{2}$} \\
Time Period ($T$)    &  Bias & RMSE &  Bias & RMSE \\ \hline
$5$   &0.0019  &0.0049  &0.0018  &0.0045 \\
$10$  &0.0004  &0.0023  &0.0008  &0.0020 \\
$30$  &0.0004  &0.0010  &0.0004  &0.0012 \\
$50$  &0.0003  &0.0008  &0.0002  &0.0008 \\
$100$ &0.0001  &0.0006  &0.0003  &0.0005 \\ \hline
\end{tabular}
\end{center}
\label{example1_4}
\end{table*}

\subsection{Setting 2}
\begin{figure*}
  \centering
  % Requires \usepackage{graphicx}
  \includegraphics[width=0.9\textwidth]{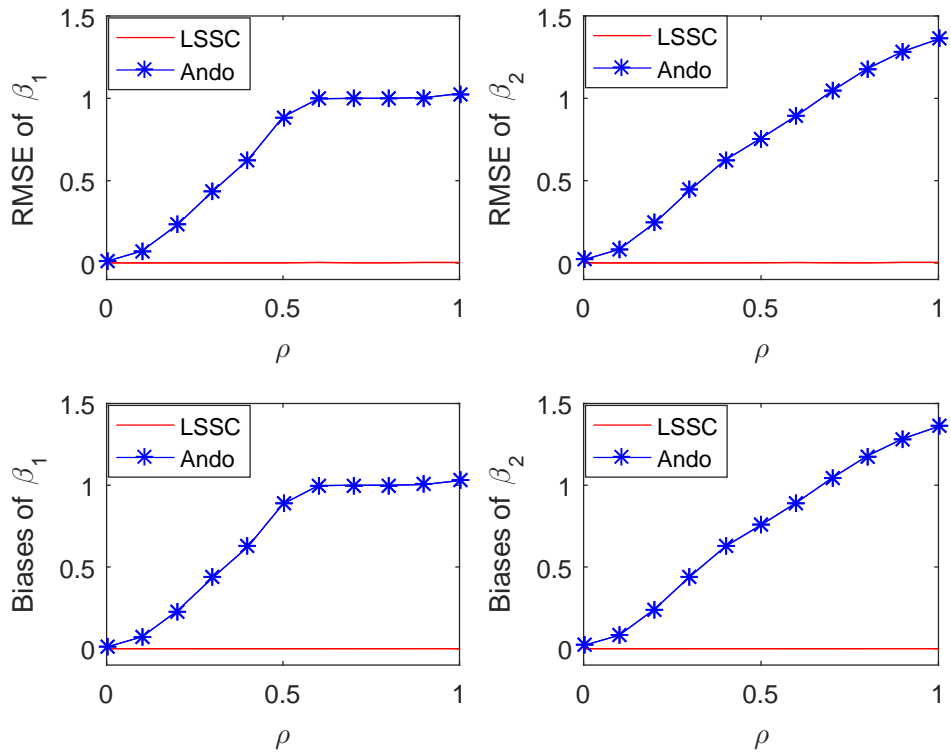}
  \caption{The simulated biases and RMSEs of the estimators of $\beta_1$ and $\beta_2$ obtained from the proposed the LSSC method and the {\it Ando-Bai} method with $\rho$ varies from $0$ to $1$ for Setting 2(b). The true values of the parameters are $\bbeta = (1, 2)$ and $N_1 = N_2 = N_3=100,T=100,r=3$. The results are based on 100 simulations for setting with $d_1=d_2=d_3=2$.}\label{compare}
\end{figure*}

The method proposed by \cite{Ando2016} is effective when the regressors are not correlated with factors and factor loadings under large $N$ and large $T$, but it does not performs well when the regressors and the factors are correlated, such as the set-up in Setting 1.
%the setup of setting 1 because of the correlation between the regressors and the factors structure.
To further compare the proposed method with the method proposed by  \cite{Ando2016}, we consider the following settings:

\begin{itemize}

\item[(a)] Let the regressors $\bx_{it}\sim Uniform(-2,2)$, and the other settings are the same as Setting 1. In this setup, the regressors are not correlated with the factors and the factor loadings.
The simulated biases and RMSEs of the estimation method proposed by \cite{Ando2016} (denoted as {\it Ando-Bai}) and the proposed LSSC for time periods $T = 10$ and 100 are presented in Tables  \ref{section6_5_1} and  \ref{section6_5_2}, respectively. From Table \ref{section6_5_1}, we can see that the proposed LSSC method has smaller biases and RMSEs compared to the {\it Ando-Bai} method when $T=10$. From Table \ref{section6_5_2}, the {\it Ando-Bai} method is performing well while the proposed LSSC method still perform better when $T = 100$.

\item[(b)] Consider the regressors and the factor loadings are correlated with the followings:
\begin{itemize}
\item the covariate $\bX$ is a $T\times N \times p$ array with
\begin{eqnarray*}
\bX_{\cdot \cdot 1}  = \rho \btau \bLambda' + \beeta_{1}
{\mbox {, }} \bX_{\cdot \cdot 2}  = \rho \btau \bLambda' + \beeta_{2},
\end{eqnarray*}
where $\bX_{\cdot \cdot 1}$ and $\bX_{\cdot \cdot 2}$ are $T \times N$ matrices;
\item $\beeta_{1} \sim Uniform_{T \times N}(-2,2)$ and $\beeta_{2} \sim Uniform_{T \times N}(-2,2)$;
\item $\btau$ is a $T\times r$ matrix of all the elements that are $1$;
\item $\rho$ is a constant which represents the correlation between the covariate $\bX$ and the factor loadings.
\end{itemize}
The other settings are the same as those settings presented in Setting 1. Figure \ref{compare} presents the simulated biases and RMSEs of the estimators for $\beta_1$ and $\beta_2$ obtained from the LSSC and {\it Ando-Bai} methods
with $\rho$ varies from $0$ to $1$. From Figure \ref{compare}, we can see that the LSSC method is more stable and gives smaller biases and RMSEs compared to the {\it Ando-Bai} method.
\end{itemize}
\begin{table*}
\caption{Simulated biases and root mean squared errors (RMSEs) of estimates based on LSSC and {\it Ando-Bai} methods for Setting 2(a). The true parameters are $\bbeta = (1, 2)$ and $N_1 = N_2 = N_3=100,T=10,r=3$. The results are based on 100 simulations for each setting.}
\centering
\label{section6_5_1}
\begin{center}
\begin{tabular}{l r r r r r r r r} \hline
&          & \multicolumn{2}{c}{{\it Ando-Bai}} &  \multicolumn{2}{c}{$LSSC$}  \\
Dimension of subspaces&    &  Bias & RMSE  &  Bias & RMSE \\ \hline
$d_1=d_2=d_3=2$&$\beta_1$  &0.1002   &0.1019    &0.0002  &0.0035  \\
               &$\beta_2$  &0.2019   &0.2027     &0.0003  &0.0037  \\
\hline
$d_1=d_2=d_3=1$&$\beta_1$  &0.1022   &0.1036   &0.0001   &0.0017 \\
               &$\beta_2$  &0.1989   &0.1997     &0.0002   &0.0020 \\
 \hline
$d_1=d_2=2,d_3=1$&$\beta_1$  &0.1006   &0.1019    &0.0063    &0.0457  \\
                 &$\beta_2$  &0.1990   &0.1999     &0.0113    &0.0491 \\
 \hline
$d_1=2,d_2=d_3=1$&$\beta_1$  &0.0990   &0.1002   &0.0040   &0.0343 \\
                 &$\beta_2$  &0.2008   &0.2015      &0.0019   &0.0367 \\
\hline
\end{tabular}
\end{center}
\end{table*}

\begin{table*}
\caption{Simulated biases and root mean squared errors (RMSEs) of estimates based on LSSC and {\it Ando-Bai} methods for Setting 2(a). The parameters are $\bbeta = (1, 2)$ and $N_1 = N_2 = N_3=100,T=100,r=3$. The results are based on 100 simulations for each setting.}
\centering
\label{section6_5_2}
\begin{center}
\begin{tabular}{l r r r r r r r r} \hline
&          & \multicolumn{2}{c}{$Ando-Bai$} &  \multicolumn{2}{c}{$LSSC$}  \\
Dimension of subspaces&    &  Bias & RMSE  &  Bias & RMSE \\ \hline
$d_1=d_2=d_3=2$&$\beta_1$  &0.0100   &0.0102    &0.0000  &0.0008  \\
               &$\beta_2$  &0.0198   &0.0199     &0.0001  &0.0010  \\
\hline
$d_1=d_2=d_3=1$&$\beta_1$  &0.0098   &0.0100  &0.0000   &0.0005 \\
               &$\beta_2$  &0.0198   &0.0199     &0.0000   &0.0005 \\
 \hline
$d_1=d_2=2,d_3=1$&$\beta_1$  &0.0097   &0.0099    &0.0003    &0.0020  \\
                 &$\beta_2$  &0.0197   &0.0198     &0.0001    &0.0013 \\
 \hline
$d_1=2,d_2=d_3=1$&$\beta_1$  &0.0099   &0.0101   &0.0003   &0.0048 \\
                 &$\beta_2$  &0.0199   &0.0200      &0.0000   &0.0025 \\
\hline
\end{tabular}
\end{center}
\end{table*}

\section{Model Selection and Possible Extensions}

In the previous sections, we assume that the number of subspaces and the dimension of subspaces are known. In this section, we extend the procedure to a more general setting in which the dimension of factors is unknown and discuss the situations that the number of subspaces for the factors and the dimension of these subspaces are unknown.

\subsection{Determine the number of subspaces for factors}

%So far we have assumed that the number of subspaces $k$ is known; however, this is not true in some situation.
One of the critical aspects of cluster analysis is to determine the number of subspaces empirically based on the observed data. For experimental data, however, there is no real ``true" number of subspaces, but only a choice of the suitable value of $k$ which can provide stable and replicable results with a good fit to the data.
In fact, the problem of estimating the subspaces number is a challenging model selection problem. Here, we are not intended to give a detailed  review of all the existing methods for obtaining the number of subspaces for factor, but we aim to provide a feasible solution based on the work by \cite{Liu_Ma2013}.

%In the following, we use the method of \cite{Liu} to illustrate the basic idea behind the determining the number of subspaces.
\cite{Liu_Ma2013} proposed a novel objective function named low-rank representation (LRR), which seeks the lowest rank representation among all the candidates that can represent the samples as linear combinations of the bases in a given dictionary. The computational procedure of LRR is to solve a nuclear norm regularized problem \citep{Fazel2002}, which is a convex optimization problem that can be solved in polynomial time.
The estimate of the number of subspaces can be obtained as \citep{Liu_Ma2013}
\begin{eqnarray}
\hat{k}=N-\text{int}\left[ \sum_{i=1}^{N}f_{\tau}(\sigma_{i}) \right],
\label{khat}
\end{eqnarray}
where $\tau$ is a cut-off threshold, $\sigma_{i}$ denotes a singular value of the normalized Laplacian matrix of the affinity matrix of data, $\text{int}[a]$ is the nearest integer of a real number $a$ and $f_{\tau}$ is a summation function which counts different values regarding that $\sigma_{i}<\tau$ defined as $f_{\tau}(\sigma)=1$ if $\sigma \geq \tau$ and $f_{\tau}(\sigma)=\log_{2}(1+\frac{\sigma^{2}}{\tau^{2}})$ if $\sigma < \tau$,
where $0<\tau<1$ is a parameter. Specifically, we can use the following steps to obtain the number of subspaces:
\begin{itemize}
\item[{\bf Step 1.}] Given $\bbeta$, update $\bF$ and $\bLambda$ by ignoring the subspace structures;
\item[{\bf Step 2.}] Given $\bF$ and $\bLambda$, update $\bbeta$;
\item[{\bf Step 3.}] Repeat Steps 1--2 until convergence occurs. Let $\by_{i}-\bx_{i}^{'}\bbeta = \bF^{'}\blambda_{i}+\bvarepsilon_{i},\;i=1,\ldots,N$, then we can compute the affinity matrix $W$ by using Algorithm 2 in \cite{Liu_Ma2013};
\item[{\bf Step 4.}] Compute the Laplacian matrix $L=I-D^{-\frac{1}{2}}WD^{-\frac{1}{2}}$, where $$D=\text{diag}(\sum_{j}[W]_{1j},\ldots,\sum_{j}[W]_{nj});$$
\item[{\bf Step 5.}] Obtain the number of subspaces, ${\hat k}$, by Eq. (\ref{khat}).
\end{itemize}

To verify the performance of the above algorithm in obtaining the number of subspaces, we assume that the numbers of subspaces in Settings 1 and 2 are unknown and
apply the above algorithm to obtain the number of subspaces based on each simulated data set. Based on the simulation study, we find that the above algorithm can obtain ${\hat k} = 3$ correctly for all the 100 simulated data sets. For future research, evaluating the performance of the above algorithm for obtaining the number of subspaces under different settings (e.g., different sample sizes, different number of subspaces, etc.) is of interest.

\subsection{Determine the dimension of factors and the dimension of subspaces}

Determination of the dimension of factors (dimension of ambient space) is an interesting research topic. The dimension of factors can be specified based on the particular practical problem using professional or expert knowledge. When professional or expert knowledge about the dimension of factors is not available, \cite{Bai_Ng2019} developed a regularization criterion to determine the number of factors, and this criterion is more stable when the nominal number of factors is inflated by the presence of weak factors or large measurement noise.
To choose the dimension of factors $r \in [0, rmax]$, the expression of the criterion is
$$\bar{r} = \min\limits_{r = 0, \cdots, rmax} \log\left[ 1-\sum\limits_{j=1}^r(D_{jj}-\gamma)^2 \right] + kg(N,T),$$
where $g_{N,T}=\frac{N+T}{NT}\log(\frac{NT}{N+T})$, $D_{jj}$ is the $j$-th singular value of the scaled observable data, and $\gamma$ is a constant threshold.
Through Monte Carlo simulation, we found that this criterion performs well in determining the number of factors when the factors and factor loadings have the subspace structure.

When the dimensions of the subspaces are unknown, determining the dimension for each subspace is still an open and challenging problem.
%In fact, it is unlikely that one model can fit any panel data perfectly. The form of the appropriate model often depends on the class of models of interest as well as the prior information available.
In this section, we suggest to obtain the solution of the optimal model selection as
\begin{eqnarray}
Z^{*} & = & \arg\min_{\mathcal{A} : \hat{Z} \subset \mathcal{A}} SSR(\hat{Z})+ \hat{\sigma}^2\sum_{j=1}^{k}d^2_j \frac{T+N_j}{NT}\log(TN_j) ,~~SSR(\hat{Z}) < \tau,  \label{model_selection}
\end{eqnarray}
where $SSR(\hat{Z})$ represents the mean squared errors under the subspaces set $\hat{Z}$ (i.e., a measure of the data fidelity), $\tau$ is the error tolerance, $d_j$ is the dimension of the $j$-th subspace and $N_j$ is the number of individuals in $j$-th subspace, $k$ is the number of subspaces, $\hat{\sigma}^2$ is the estimated variance, and $\sum_{i=1}^{k}d^2_i \frac{T+N_i}{NT}\log(TN_i)$ is the penalty term which measures the model complexity under the subspaces set $\hat{Z}$. The proposed criterion can be viewed as a tradeoff between how well the model fits the data and the model complexity. It can be shown that the penalty function $d^2_j \frac{T+N_j}{NT}\log(TN_j)\to 0$ and $\min\{N,T\}d^2_j \frac{T+N_j}{NT}\log(TN_j)\to \infty$ as $T,N\to \infty$ and $T/N$ converges to constant.

\begin{theorem}{Theorem 5.}{} Suppose that Assumptions A--F hold and $T/N \rightarrow \rho>0$, then
the dimensions of subspaces $\{\hat{d}_1,\cdots,\hat{d}_k\}$ obtained by using Eq. (\ref{model_selection}) converge in probability to the true dimensions of subspaces $\{d_1^0, \cdots, d_k^0\}$.
\label{criterion}
\end{theorem}

To examine the proposed method for model selection, we simulated the panel data from the models with number of factors $r = 3$ and 4 and then obtain the solution of the optimal model selection in Eq. \eqref{model_selection} with different number of units in each subspace and different time period $T$. %In order to avoid model fitting failure, we take the error tolerance $\tau=\frac{10 (d_1^3 + \cdots + d_{k-1}^3)}{\min\{N,T\}}$.
We consider that there is no covariate, i.e., $\beta = 0$ and we use the error tolerance $\tau=\frac{10 (d_1^3 + \cdots + d_{k-1}^3)}{\min\{N,T\}}$.
The simulated percentages of identifying the correct dimension of subspaces (based on 1000 simulations for each setting) are presented in Tables \ref{model_selection1} and \ref{model_selection2} for $r = 3$ and 4 with three and four subspaces, respectively.
From the simulation results in Tables \ref{model_selection1} and \ref{model_selection2}, the proposed model selection method performs reasonably well in the case of hyperplane, i.e., these subspaces have the same dimensions. Compared with the case of the hyperplane, when the dimensions of the subspaces are not all the same, the
simulated percentages of identifying the correct dimension can be lower to about 80\%.

\begin{table*}
  \caption{Simulated percentages of identifying the correct dimension of subspaces using Eq. \eqref{model_selection} with different $(N,T)$ and $N_1=N_2=N_3=N/3$, and the number of factors is $r=3$.}
  \centering
  \label{model_selection1}
 \begin{center}
%Compared our proposed methods with the randomized exchange algorithm (REX) for $D$-optimality.
\begin{tabular}{l r r r r r r r} \hline
                            & \multicolumn{4}{c}{$(N, T)$} \\
 Dimension of subspaces     &    (600, 300)      & (600, 600)       & (1500, 300)         & (1500, 600)  \\\hline
 $d_1=d_2=d_3=1$            &97.10      \%      &100.00 \%        &96.70  \%           &100.00 \%  \\
 $d_1=d_2=d_3=2$            &100.00      \%      &100.00 \%       &100.00 \%           &100.00 \%  \\
 $d_1=d_2=2,d_3=1$          &79.20      \%      &80.40 \%         &87.00  \%           &93.70 \%  \\
 $d_1=d_2=1,d_3=2$          &90.20      \%      &80.10 \%         &96.70  \%           &100.00 \%  \\   \hline
\end{tabular}
\end{center}
\end{table*}

\begin{table*}
   \caption{Simulated percentages of identifying the correct dimension of subspaces using Eq. \eqref{model_selection} with different $(N,T)$ and $N_1=N_2=N_3=N_4=N/4$, and the number of factors is $r=4$.}
  \centering
 \label{model_selection2}
 \begin{center}
%Compared our proposed methods with the randomized exchange algorithm (REX) for $D$-optimality.
\begin{tabular}{l r r r r r} \hline
& \multicolumn{4}{c}{$(N, T)$} \\
 Dimension of subspaces     & (800, 300)              & (800, 800)        & (2000, 300)             & (2000, 800)        \\\hline
 $d_1=d_2=d_3=d_4=1$        &95.00 \%                &100.00\%       &92.50 \%                &100.00 \%              \\
 $d_1=d_2=d_3=d_4=2$        &99.00 \%                &99.90\%        &99.40 \%                &100.00 \%             \\
 $d_1=d_2=d_3=d_4=3$        &100.00 \%               &100.00\%       &100.00 \%               &100.00 \%             \\
 $d_1=d_2=d_3=2,d_4=1$      &85.60 \%                &84.10\%        &85.40 \%                &84.40 \%           \\\hline
% $d_1=3,d_2=d_3=d_4=2$      &58.76 \%                &54.08\%        &67.96 \%                &69.49 \%             \\ \hline
\end{tabular}
 \end{center}
\end{table*}

\section{Real Data Application}

In this section, we illustrate the proposed methodologies by using the real data provided by \cite{Bonhomme2015} and studying the linkage between income growth and democracy across different countries. Following \cite{Bonhomme2015}, we use the linear dynamic model to identify the group membership and the linkage between income growth and democracy across countries, i.e.,
\begin{eqnarray*}
democracy_{it}&=&\theta_1 democracy_{i(t-1)} + \theta_2log GDPpc_{i(t-1)} + \blambda_{g_i,i} \bff_{g_i,t}+v_{it},
\label{application}
\end{eqnarray*}
where
$democracy_{it}$ is the democracy index (measured by the Freedom House indicator with values in between 0 (the lowest) and 1 (the highest)) for the $i$-th country at time $t$, $GDPpc_{it}$ is the GDP per capita of the $i$-th country at time period $t$, and $\blambda_{g_i,i}$ and $\bff_{g_i,t}$ are the unobservable grouped factor loadings and factors, respectively.
Here, the dependent variable is the country's democracy index and the explanatory variables are the first-order lagged democracy index and the income of a country measured by the logarithm of GDP per capita.

The data set contains a balanced panel of 90 countries and 7 periods at a five-year interval over 1970--2000.
First, using the information criteria suggested in \cite{Bai_Ng2019} to estimate the number of factors, we obtain the dimension of factor space as $r = 5$. Then, the number of subspaces is estimated as $k = 3$ based on Eq. (\ref{khat}). The results are consistent with those presented in \cite{SuPhillips2016}. Next, we use the criterion in Eq. (\ref{model_selection}) to select the optimal model, and the results show that the optimal model have the dimensions $d_1 = d_2 = d_3 = 4$. Finally, we use BAI, GFE and LSSC methods to obtain the parameter estimates as $(\hat{\theta}_1, \hat{\theta}_2)$ and corresponding fitting errors (defined as $\hat{SSR}=\frac{1}{NT}\sum\limits_{i=1}^N\sum\limits_{t=1}^T(democracy_{it}-\theta_1 democracy_{i(t-1)}
-\theta_2log GDPpc_{i(t-1)} - \blambda_{g_i,i} \bff_{g_i,t})^2$). The estimated results are presented in Table \ref{empirical_study}. From Table \ref{empirical_study}, we can see that all these estimates imply the effect of income on democracy is positive, but the LSSC method has the smallest fitting error.

\begin{table*}
   \caption{
   BAI, GFE and LSSC methods are used to obtain the parameter estimates as $(\hat{\theta}_1, \hat{\theta}_2)$ and corresponding fitting errors $\hat{SSR}$, where the number of factors is $r=5$ and the number of groups is $k=3$.}
  \centering
 \label{empirical_study}
 \begin{center}
%Compared our proposed methods with the randomized exchange algorithm (REX) for $D$-optimality.
\begin{tabular}{l r r r r} \hline
 Methods                  & $(\hat{\theta}_1, \hat{\theta}_2)$  & $\hat{SSR}$  \\\hline
BAI                 & (0.6023,0.3729)              &0.0024  \\
GFE                 & (0.0869, 0.1723)             &0.1823  \\
LSSC                & (0.8330, 0.3540)             &3.5714e-04   \\ \hline
\end{tabular}
\end{center}
\end{table*}

\begin{figure*}
  \centering
  % Requires \usepackage{graphicx}
  \includegraphics[width=0.9\textwidth]{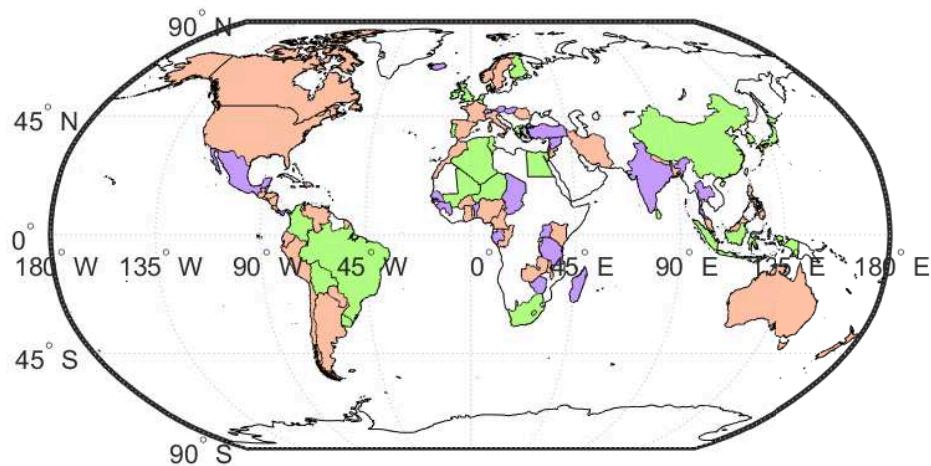}
  \caption{Subspaces clustering for 90 countries}
  \label{map1}
\end{figure*}

In order to visualize the group membership obtained by the proposed method, we put these grouped countries on a world map in Figure \ref{map1} in which the countries in the same group are represented in the same color.
The detailed lists of grouped countries are presented as followings:
\begin{itemize}
\item Group 1 (45 countries): Argentina, Australia, Bangladesh, Burkina Faso, Burundi, Cameroon, Canada, Chile, Congo, Costa Rica, Denmark, Dominican Rep., Ecuador, El Salvador, France, Gambia, Ghana, Guatemala, Honduras, Iran, Israel, Italy, Jamaica, Jordan, Kenya, Luxembourg, Malawi, Malaysia, Morocco, Nepal, New Zealand, Nicaragua, Nigeria, Norway, Paraguay, Peru, Philippines, Romania, Spain, Sweden, Togo, Trinidad and Tobago, United States, Venezuela, Zambia.
\item Group 2 (24 countries): Algeria, Belgium, Bolivia, Brazil, China, Colombia, Egypt, Finland, Greece, Indonesia, Ireland, Japan, Korea, Lesotho, Mali, Netherlands, Niger, Portugal, Rwanda, South Africa, Sri Lanka, Tunisia, United Kingdom, Uruguay.
\item Group 3 (21 countries): Austria, Barbados, Benin, Chad, Gabon, Guinea, Hungary, Iceland, India, Madagascar, Mauritius, Mexico, Panama, Senegal, Switzerland, Syria, Tanzania, Thailand, Turkey, Uganda, Zimbabwe.
\end{itemize}

From these groupings, it can be seen that most of the early developed countries are distributed in the first group, which has a certain relationship with the economic and political structure. It includes the United States and Canada, most of the countries in continental Europe, coastal countries of South America and Australia. Most of the countries in the second group are developing countries with rapid economic development in Asia, Africa and South America, which includes China, Brazil and South Africa. Japan and South Korea also belong to the second group because they are both countries with high-speed economic development during this period and similar culture and policies. Most of the countries in the third group have slower development and relatively backward economies and policies during this period.

\section{Concluding Remarks}

In this paper, we consider a panel data model that allows the covariates and the unobservable latent variables to be correlated.
% and the relationship between group membership and observed covariates is left unrestricted.
We propose a subspace clustering method for factor loadings of the panel data model that captures the grouped unobserved heterogeneity.
The common regression parameters, grouped unobservable factor structure and group membership can be estimated simultaneously with the proposed method.
The asymptotic results show that the subspace clustering and the estimators are consistent.
The Monte Carlo simulation results show that the proposed methodologies outperform the existing methods under different settings.
Under the model considered in this paper, we propose a consistent model selection criterion to determine a suitable subspace dimension.
We also discuss some possible future research directions in determining the number of subspaces and factor dimension when these values are unknown.
These issues are under investigation and we hope to report the results in a future paper.
%%%%%%%%%%%%%%%%%%%%%%%%%%%%%%%%%%%%%%%%%%%%%%%%%%%%%%%%%%%%%%%%%%%%%%%%%%%%%%%%%%%%%%%%%%%%%%%%%%%%%%%%%%%%%%%%%%%%%%%%%%%%
%\section*{Acknowledgments}

%The authors thank the Associate Editor and the referees for their helpful comments and suggestions which improved the presentation and the quality of this paper. H. K. T. Ng's research work was supported by a grant from the Simons Foundation (\#709773 to Tony Ng).
%%%%%%%%%%%%%%%%%%%%%%%%%%%%%%%%%%%%%%%%%%%%%%%%%%%%%%%%%%%%%%%%%%%%%%%%%%%%%%%%%%%%%%%%%%%%%%%%%%%%%%%%%%%%%%%%%%%%%%%%%%%%

\end{document}